\documentclass{./aa}
\usepackage{graphics}
\usepackage{epsfig}

\begin{document}

   \title{Circular Polarization of Water Masers in the Circumstellar
   Envelopes of Late Type Stars}
   \titlerunning{Circular Polarization of Circumstellar Water Masers}


   \author{W.H.T. Vlemmings\inst{1}\and
        P.J. Diamond\inst{2}\and
        H.J. van Langevelde\inst{3}\
          }

   \offprints{WV (vlemming@strw.leidenuniv.nl)}

   \institute{Sterrewacht Leiden, Postbus 9513, 2300 RA Leiden, 
              the Netherlands
        \and
         Jodrell Bank Observatory, University of Manchester, Macclesfield,
                    Cheshire, SK11 9DL, England     
        \and
        Joint Institute for VLBI in Europe, Postbus 2, 
                7990~AA Dwingeloo, The Netherlands
        }

   \date{Received ; accepted }

   \abstract{ We present circular polarization measurements of
     circumstellar H$_{2}$O masers. The magnetic fields in
     circumstellar envelopes are generally examined by polarization
     observations of SiO and OH masers. SiO masers probe the high
     temperature and density regime close to the central star. OH
     masers are found at much lower densities and temperatures,
     generally much further out in the circumstellar envelope.  The
     circular polarization detected in the (6$_{16}$--5$_{23}$)
     rotational transition of the H$_{2}$O maser can be attributed to
     Zeeman splitting in the intermediate temperature and density
     regime.  The magnetic fields are derived using a general, LTE
     Zeeman analysis as well as a full radiative transfer method
     (non-LTE), which includes a treatment of all hyperfine components
     simultaneously as well as the effects of saturation and unequal
     populations of the magnetic substates. The differences and
     relevances of these interpretations are discussed extensively. We
     also address a non-Zeeman interpretation as the cause for the
     circular polarization, but this is found to be unlikely. We favor
     the non-LTE analysis. The H$_2$O masers are shown to be
     unsaturated, on the basis of their line widths and the lack of
     linear polarization. The field strengths are compared with
     previous detections of the magnetic field on the SiO and OH
     masers. Assuming a $r^{-2}$ dependence of the magnetic field on
     the distance to the star, similar to a solar-type magnetic field,
     our results seem to indicate that we are probing the highest
     density maser clumps at the inner edge of the H$_2$O maser
     region. This allows us to estimate the density of the clumps, and
     the extent of the H$_2$O maser region. We show that the
     magnetic pressure dominates the thermal pressure by a factor of
     $20$ or more. We also give an order of magnitude estimate of the
     magnetic field on the surface of the stars.  In particular we
     discuss the differences between Supergiants and Mira variable
     stars.   }

\maketitle

\begin{figure} 
   \resizebox{0.9\hsize}{!}{\includegraphics{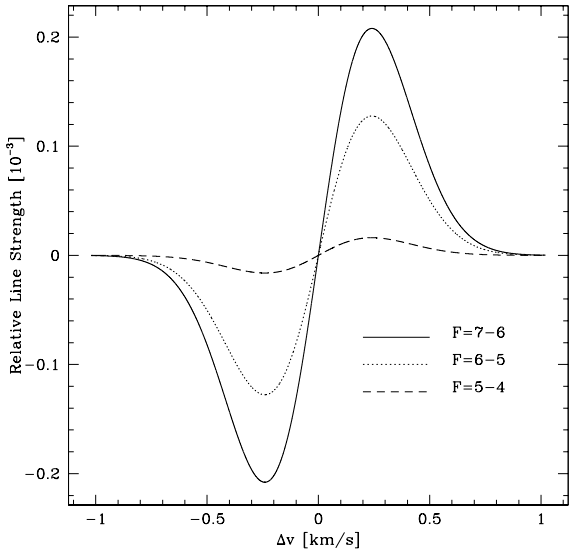}} \hfill \caption{
   Synthetic V-spectra for an external field of 50 mG and a line width
   of $\Delta v_{\rm L}=0.8$ km~s$^{-1}$.}  \label{vs}
\end{figure}
\begin{figure} 
   \resizebox{0.9\hsize}{!}{\includegraphics{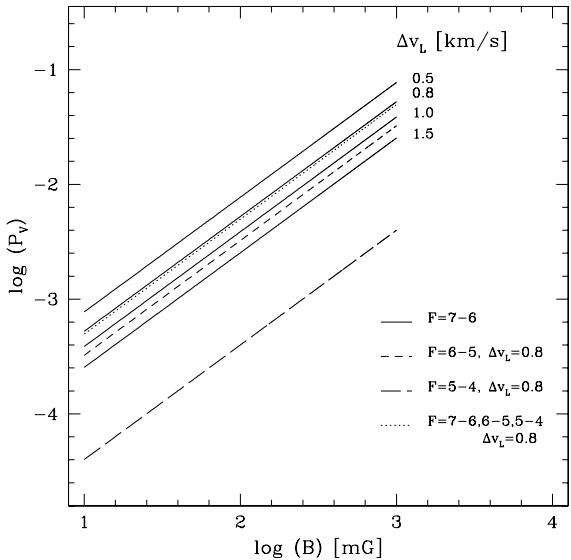}} \hfill
   \caption{$B-P_{\rm V}$ relations. The solid lines represent the
   $F=7-6$ transition for different $\Delta v_{\rm L}$. The short- and
   long-dashed lines are the $F=6-5$ and the $F=5-4$ transition
   respectively. The dotted line is a typical best fitted model
   including all three hyperfine lines.} \label{bv}
\end{figure}

\section{Introduction}
\label{intro} 
A large majority of stars go through a period of high mass loss at the
end of their evolution while climbing the asymptotic giant branch
(AGB). This mass loss, of the order of $10^{-7}$ to $10^{-4} M_\odot
/{\rm yr}$, is the main source for replenishing interstellar space
with processed materials. Thus the mass loss mechanism is an important
subject of study. In late type stars, this high mass loss produces a
circumstellar envelope (CSE) in which several maser species can be
found. These masers, especially SiO, H$_{2}$O and OH, are excellent
tracers of the kinematics of the CSEs. The role of magnetic fields in
the mass loss and formations of CSEs is still unclear. Polarization
observations of circumstellar masers can reveal the strength and
structure of the magnetic field throughout the CSE.  Observations of
SiO maser polarization have shown highly ordered magnetic fields close
to the central star, at radii of 5-10 AU where the SiO maser emission
occurs (e.g. Kemball \& Diamond, 1997). The measured circular
polarization indicates magnetic field strengths of $\approx 5-10$ G,
when assuming a standard Zeeman interpretation. However, a non-Zeeman
interpretation has been proposed by Wiebe \& Watson (1998), which only
requires fields of $\approx$~30 mG.  At much lower densities and
temperatures and generally much further from the star, OH maser
observations measure fields of $\approx 1$ mG (e.g. Szymczak \& Cohen,
1997; Masheder et al., 1999) and there is little dispute of the Zeeman
origin of the polarization. Until recently, no information on the
magnetic fields at distances of a few hundred AU from the star was
available.  This is the region where the H$_{2}$O maser emission
occurs. Because water is a non-paramagnetic molecule, determination of
the magnetic field is significantly more difficult. The fields
expected in the H$_{2}$O maser region are stronger than the fields
measured for the OH masers; since they probably occur in gas that is a
factor of $10-1000$ more dense and closer to the central star
than the OH masers. So we expect fields of a few tens to a few
hundred mG.  For these fields the Zeeman splitting of H$_{2}$O is
extremely small, only $\approx~10^{-3}$ times the typical half-power
width of the H$_{2}$O maser line ($\Delta\nu_{\rm L} \approx 30$ kHz).
However, Vlemmings et al. 2001 (hereafter V01) have shown that in the
presence of such magnetic fields the Zeeman splitting can be detected
with high spectral resolution polarization observations. They
presented the first results of circular polarization measurements of
circumstellar H$_2$O masers on the masers around the supergiant star
S~Per. A 'basic', LTE, Zeeman analysis was used to infer the magnetic
field strength along the line of sight. This analysis did not include
the interaction between the various hyperfine components of the 22~GHz
H$_2$O transition. The observations confirmed the expected field
strength, finding a field of $\approx 250$ mG. The method used in V01
was similar to the method used by Fiebig \& G\"usten (1989, hereafter
FG) to analyze the circular polarization of strong interstellar H$_2$O
maser features.  Here we present observations for a sample of 4 stars.
We discuss the data reduction path, and in addition to the FG method,
we fit our observation to the theoretical, non-LTE, models of Nedoluha
\& Watson (1992), hereafter NW. These include hyperfine interaction as
well as saturation effect. The non-Zeeman interpretations presented in
Nedoluha \& Watson (1990) and Wiebe \& Watson (1998) are also
discussed. Finally we present measurements of the linear polarization.

\section{Theoretical Framework}
\subsection{Standard Zeeman interpretation}

\begin{figure*}
   \resizebox{\hsize}{!}{\includegraphics{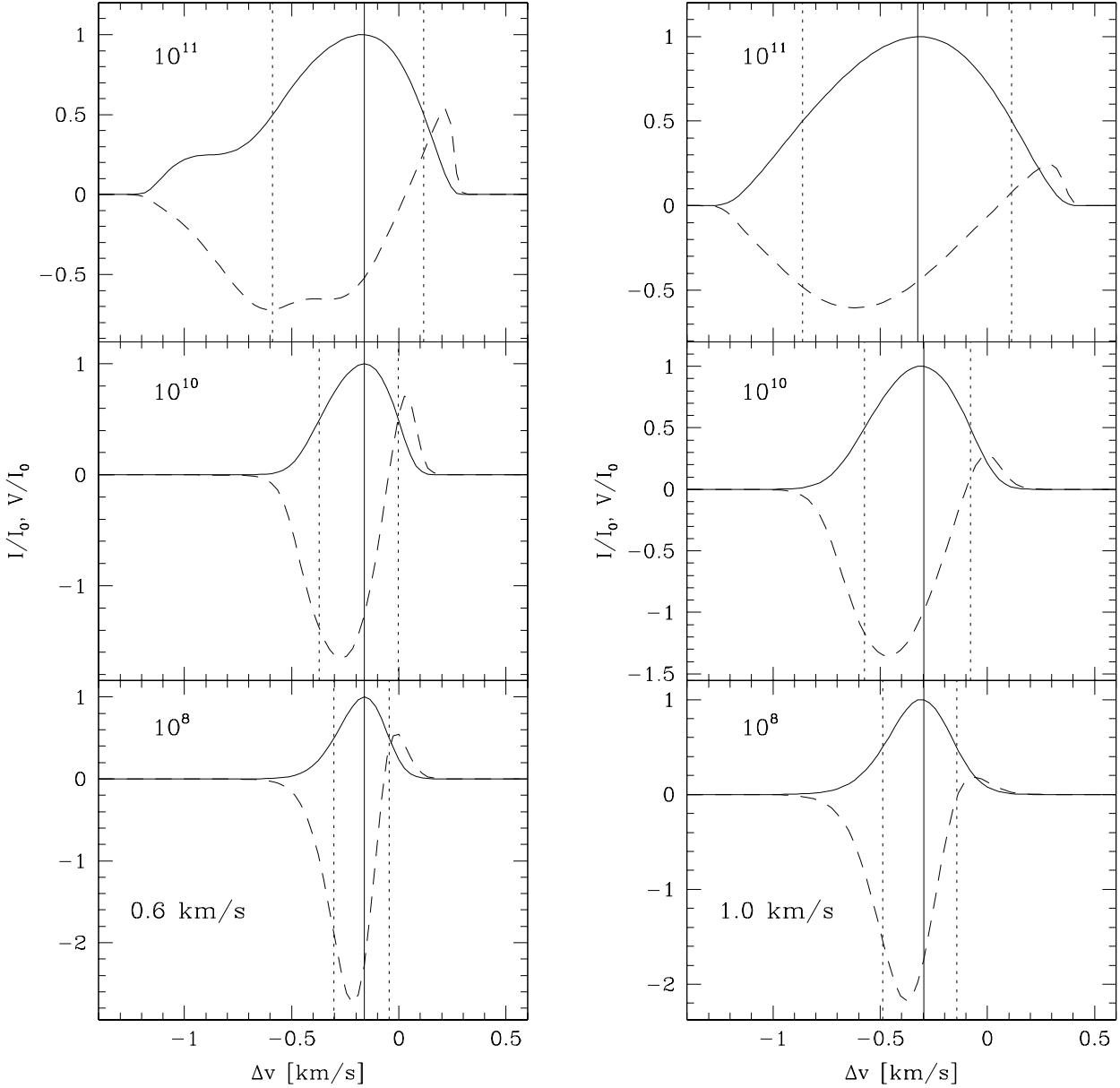}} \hfill
   \caption{Fractional circular polarization in units of
   $0.1 \%$. The solid line is the total intensity Stokes $I$, the
   long-dashed line is Stokes V. Left has $v_{\rm thermal} = 0.6$ km/s,
   right has $1.0$ km/s. The models are labeled for emerging
   brightness temperature $T_{\rm b}\Delta\Omega = 10^{11}, 10^{10}$
   and $10^8$ from top to bottom. The solid vertical line indicates
   the peak and the short dashed lines the FWHM of the profile.}
\label{nwf}
\end{figure*}
\begin{figure*} 
   \resizebox{\hsize}{!}{\includegraphics{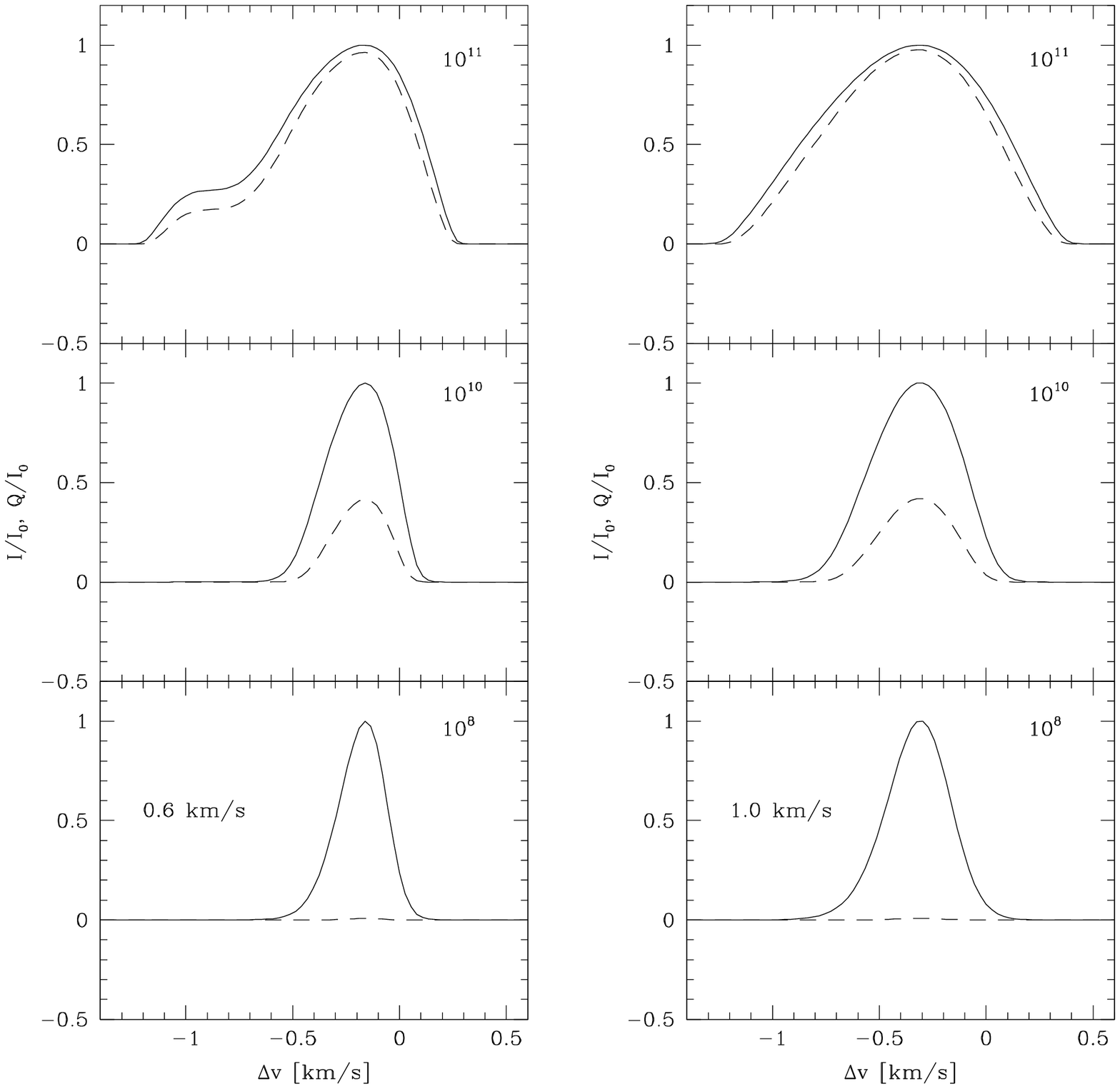}} \hfill
   \caption{Linear polarization for $\theta = 75^\circ$. The dashed
   line is linear polarization is given in units of $10 \%$. Labelling
   is similar as Fig.~\ref{nwf}} \label{lin}
\end{figure*}

\subsubsection{Basic LTE analysis}
 Our first treatment of the standard Zeeman interpretation closely
follows the analysis performed in FG and is more thoroughly discussed
in both V01 and FG. This is a standard method that assumes LTE. Thus,
the narrowing and rebroadening of the lines as a result of maser
saturation are not reproduced. We have included the possibility of
multiple masing hyperfine components, assuming the $F=7-6$ hyperfine
component to be dominant.  Each hyperfine component will split into 3
groups of lines ($\sigma^{+}, \sigma^{-}$ and $\pi$). For a magnetic
field B parallel to the line of sight, the Zeeman pattern consists of
the two circular polarized $\sigma$ components only. The right- and
left-handed (RR and LL) spectra, corresponding to the $\sigma^{\pm}$
components will only be slightly shifted against each other. This
corresponds to a characteristic frequency shift of the order of
$\Delta \nu_{\rm Z} \approx 10^3$ Hz$\cdot B_{\rm [Gauss]}$.  As a
result, the observed V-spectrum (RR-LL) will be a sine-shaped
function, corresponding to the derivative $I'$ of the total power
spectrum. The amplitude of this function depends on the maser line
width, the magnetic field strength, and on which hyperfine components
actually contribute to the maser. By calculating synthetic V-spectra
we find the following relation for the percentage of circular
polarization:
\begin{eqnarray}
P_{\rm V} & = & (V_{\rm max} - V_{\rm min})/I_{\rm max} \nonumber\\
& = & 2\cdot A_{\rm F-F'}\cdot B_{\rm [Gauss]} \rm{cos}\theta/\Delta v_{\rm L}[\rm{km~s}^{-1}]. 
\label{eq2}
\end{eqnarray}
Here $V_{\rm max}$ and $V_{\rm min}$ are the maximum and minimum of the
synthetic V-spectrum fitted to the observations. $\Delta v_{\rm L}$ is
the full width half-maximum (FWHM) of the total power spectrum, and
$I_{\rm max}$ is the peak flux of the maser feature. $B$ is the
magnetic field strength and $\theta$ the angle between the magnetic
field lines and the line of sight.  The $A_{\rm F-F'}$ coefficient
depends on the masing hyperfine components. The $A_{\rm F-F'}$
coefficients have been determined by calculating $P_{\rm V}$ from
synthetic V-spectra, determined for a series of magnetic field
strengths $B$, and for the different hyperfine components.  Some examples
of synthetic V-spectra for the three hyperfine lines are shown in
Fig.~\ref{vs}.  The $B-P_{\rm V}$ relation is shown in Fig.~\ref{bv}
for different hyperfine components and line widths.  For the different
hyperfine transitions separately, FG find $A_{\rm F-F'} = 0.013, 0.08$
and $0.01$ for the $7-6, 6-5$ and $5-4$ transitions respectively. For
the fitted combinations of hyperfine components we generally find
$A_{\rm F-F'} \approx 0.011$.

\subsubsection{Full radiative transfer, non-LTE analysis}

\begin{figure*}
   \resizebox{\hsize}{!}{\includegraphics{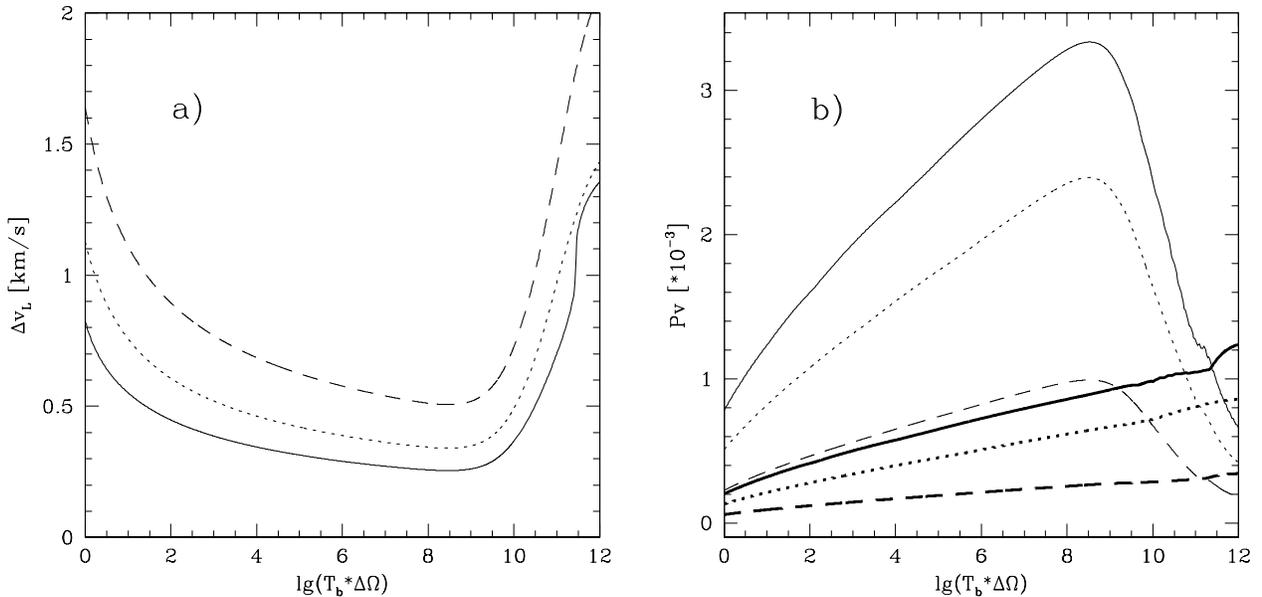}} \hfill
   \caption{a) Spectral line width versus emerging brightness
   temperature $T_{\rm b}\Delta\Omega$. The solid, dotted and dashed
   lines have $\Delta v_{\rm thermal} = 0.6, 1.0$ and $2.0$ km/s
   respectively. b) $P_{\rm V}$ versus $T_{\rm
   b}\Delta\Omega$. Labeling is similar to a), thin lines have
   $\theta = 0^\circ$, thick lines correspond to $\theta = 75^\circ$.}
\label{lwdth}
\end{figure*}

The transfer equations for the polarized radiation of the 22 GHz
H$_2$O masers in the presence of a magnetic field were solved in
NW. We have redone the calculations to fit the theoretical total
intensity and circular polarization spectra to our observations.  They
were performed for the combination of the three strongest hyperfine
components which have previously been verified to make the only
significant contribution (Deguchi \& Watson, 1986; Nedoluha \& Watson,
1991). We solve a related set of 99 equations of state for all the
magnetic substates of the relevant hyperfine components.

The goal of NW was to asses the validity of Eq.~\ref{eq2} when
specific effects of maser radiation are included in the analysis.  The
LTE treatment described above for example, does not include: narrowing
and rebroadening of the maser line with increasing optical depth,
hyperfine interaction, unequal populations of the magnetic substates
and the generation of circular polarization due to changes in
direction of the linear polarization. To solve the equations of state,
some assumptions have been made. The Zeeman frequency shift $g\Omega$,
is assumed to be much greater than the rate for stimulated emission
$R$, decay rate $\Gamma$ and cross-relaxation rate $\Gamma_\nu$. In
this regime the off-diagonal elements of the density matrix which
describes the molecular states are negligible, which greatly
simplifies the calculations. The maser is assumed to propagate nearly
one-dimensionally. Additionally, the pumping has a Maxwellian
velocity dependence and is the same for all magnetic substates.

For expected magnetic fields of $\approx$ 100 mG the values of
$g\Omega$ for the stronger hyperfine transitions are about 1000~${\rm
s}^{-1}$. The rate for stimulated emission can be estimated by:

\begin{equation}
R \simeq AkT_{\rm b}\Delta\Omega/4\pi\hbar\omega
\label{eq3}
\end{equation}

Here $T_{\rm b}$ is the brightness temperature, $\Delta\Omega$ the
beaming solid angle for the maser radiation and $A$ the Einstein
coefficient. For $T_{\rm b}\Delta\Omega~=~10^{12}$, $R\approx
100~{\rm s}^{-1}$. Thus $g\Omega\gg R$, required for ignoring the
off-diagonal elements of the density matrix, is satisfied for most of
the relevant brightness regime, up to $T_{\rm b}\Delta\Omega =
10^{12}$. For higher brightness temperatures the results for
$\theta\neq 0^\circ$ may be unreliable. For $\theta=0^\circ$ the
off-diagonal elements are not present, thus the results are
reliable up to the highest values of $T_{\rm b}\Delta\Omega $.

Detailed calculations of the cross-relaxation rate ($\Gamma_\nu$),
which is expected to be considerably larger than the decay rate
($\Gamma$), give $\Gamma_\nu = 2~{\rm s}^{-1}$ at temperatures of
400~K (Anderson \& Watson, 1993). So $g\Omega\gg \Gamma_\nu \gg \Gamma$
is satisfied as well. The calculations here use $(\Gamma+\Gamma_\nu) =
1~{\rm s}^{-1}$, but NW have shown that the resulting intensities
scale with the decay rate and the cross-relaxation rate as $[{\rm
flux}/(\Gamma+\Gamma_\nu)]$.

 To perform the numerical calculations, we had to chose the strength
of the radiation incident onto the masing region. The calculations
presented here are based on a radiation of $T_{\rm b}\Delta\Omega =
0.1$~K~sr. This continuum radiation is taken to be unpolarized. It was
verified in NW that the results are insensitive to the chosen value,
and we have confirmed this result.

 In contrast to the LTE analysis presented above, the results
from the non-LTE transfer treatment are V-spectra which are not
anti-symmetric. This can be seen comparing the non-LTE spectra in
Fig.~\ref{nwf} to the synthetic LTE spectra in Fig.~\ref{vs}. The
spectra are not proportional to the derivative of the total power
intensity profile, I', as used before, due to the blending of
hyperfine lines with different $g\Omega$.

Fig.~\ref{nwf} also shows that for high $T_{\rm b}\Delta\Omega$, the
total intensity spectra are increasingly less Gaussian in shape. The
line shape thus enables us to estimate the saturation level of the
maser. The model with $T_{\rm b}\Delta\Omega=10^8$ corresponds to a
completely unsaturated maser. The models with $T_{\rm b}\Delta\Omega =
10^{10}$ and $10^{11}$ correspond to slightly and fully saturated
masers respectively. The non-LTE analysis also produces linear
polarization for $\theta \ne 0^\circ$. Stokes $Q$ is shown in
Fig.~\ref{lin}, which indicates that for $\theta=75^{\circ}$, linear
polarization of up to $\approx 10\%$ is observed for $T_{\rm
b}\Delta\Omega \approx 10^{11}$.  NW show that for slightly saturated
masers, linear polarization of a few percent should be observed when
$\theta > 15^{\circ}$.

The calculations have been performed for several different values of
intrinsic thermal width ($v_{\rm thermal}$). This is the FWHM of the
Maxwellian distribution of particle velocities. Assuming a kinetic
temperature of $T$~K, $v_{\rm thermal} \approx 0.5(T/100)^{1/2}$ (NW).

The variation of line width with brightness temperature is shown in
Fig.~\ref{lwdth}a for different values of $v_{\rm thermal}$. This
relation does not depend on the angle $\theta$.  We see a gradual
narrowing of the maser line, and rebroadening occurs when the maser
starts to become saturated, at $T_{\rm b}\Delta\Omega \approx
10^{10}$.  The relation between the fractional circular polarization
$P_{\rm V}$ and $T_{\rm b}$ does depend on $\theta$ and is shown for
different values of $\theta$ and $v_{\rm thermal}$ in
Fig.\ref{lwdth}b.  The fractional circular polarization decreases with
increasing maser intensity as the total intensity grows faster than
the circular polarization. Both these relation scale with
$[{\rm flux}/(\Gamma+\Gamma_\nu)]$, similar to the intensity profiles.

\begin{figure}
   \resizebox{\hsize}{!}{\includegraphics{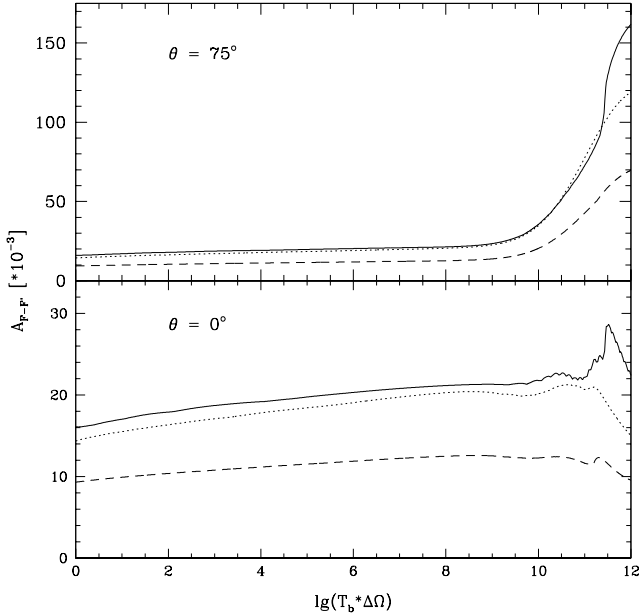}} \hfill
   \caption{$A_{F-F'}$ coefficients as a function of $T_{\rm
   b}\Delta\Omega$. The lower panel shows $\theta = 0^\circ$, the
   upper panel $\theta = 75^\circ$. The solid, dotted and dashed lines
   correspond to $\Delta v_{\rm thermal} = 0.6, 1.0$ and $2.0$ km/s
   respectively.}
\label{Aff2}
\end{figure}

From these relations we can now determine the coefficients $A_{\rm
  F-F'}$ from Eq.\ref{eq2} which are shown in Fig.\ref{Aff2}. As can
  be seen, $A_{\rm F-F'}$ depends strongly on $v_{\rm thermal}$. For
  small angles $\theta$ there is no strong dependence on $T_{\rm b}$,
  but for $\theta > 45^\circ$, $A_{\rm F-F'}$ can change by over a
  factor of $6$ for high $T_{\rm b}$.  Neufeld \& Melnick (1990) argue
  that the low values of $v_{\rm thermal}$ can be eliminated, on the
  basis of realistic H$_2$O maser pumping schemes.  Values as high as
  $v_{\rm thermal} \approx 2.0$ km/s indicate temperatures of more
  than $1000$ K, which is expected to be too high according to the
  same analysis by Neufeld \& Melnick. From linewidth analysis we
  conclude that $v_{\rm thermal} \approx 1.0$~km/s is the most
  realistic estimate, which yields $A_{\rm F-F'} \approx 0.018$.  This
  value does not decrease by more than $10~\%$ if we take $v_{\rm
  thermal} \approx 1.5$ corresponding to the upper temperature limit
  of $1000$~K.

\begin{figure}
   \resizebox{\hsize}{!}{\includegraphics{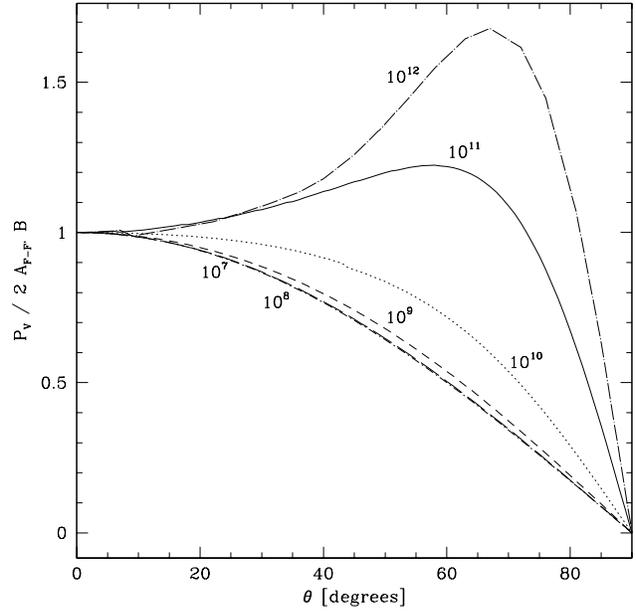}}
   \hfill \caption{$\theta$-dependence of Eq.~\ref{eq2} for increasing
   emerging brightness temperature $T_{\rm b}\Delta\Omega$. The lines
   for $T_{\rm b}\Delta\Omega = 10^{7}$ and $10^8$ coincide and are
   the same as the lines for lower brightness temperatures.}
\label{theta}
\end{figure}

The effect of the angle $\theta$ has recently been examined by Watson
\& Wyld (2001) for the general case of a transition with low angular
momentum (J=1-0 and J=2-1). They conclude that for masers that are at
least partially saturated the circular polarization does not decrease
with increasing $\theta$ until the magnetic field and the propagation
direction are nearly perpendicular. We have performed similar
calculations for the specific 22 GHz H$_2$O maser transitions and the
results are shown in Fig.~\ref{theta}, which confirm for transitions
with higher angular momentum, the analysis of Watson \& Wyld.

\subsection{Non-Zeeman interpretation}

Circular polarization can also occur due to the change of the axis of
symmetry for the molecular quantum states from parallel to the
magnetic field, to parallel to the direction of propagation. When
$g\Omega \gg R$, the direction of the magnetic field is the axis of
symmetry. When $g\Omega \ll R$, the molecules interact more strongly
with the radiation than with the magnetic field, and the direction of
propagation is the axis of symmetry. Initially $g\Omega \gg R$, but as
the maser propagates through the medium and becomes more intense,
$g\Omega \ll R$ could possibly be satisfied. This was discussed by
Nedoluha \& Watson (1990) and was shown to lead to an overestimation
of the magnetic field by up to a factor of 1000. However, the study in
NW, and the line width observed here, show that the effect described
above is unimportant for H$_2$O masers. The line widths show that in
almost all cases $g\Omega \gg R$ is satisfied.

Wiebe \& Watson (1998) have shown that the propagation of linear
polarization can also result in circular polarization. For linear
polarizations of $\approx 10\%$, the resulting circular polarization
is of the same order of magnitude as the polarization due to the
regular Zeeman interpretation, while magnetic fields could again be a
factor of 1000 less. Below we indicate that also this non-Zeeman
interpretation is unlikely due to the lack of linear polarization.

\subsection{Fitting Method}
\label{sect}

\subsubsection{LTE Zeeman}

In the case of the LTE Zeeman models we first fit a combination of the
3 strongest hyperfine components to the total power spectrum.  From
this we get the line width $\Delta v_{\rm L}$ of the maser feature and
the contribution to the line of the hyperfine components.  For this
combination we calculate a synthetic V-spectrum, such as shown in
Fig.~\ref{vs} for the separate hyperfine components. The synthetic
spectrum is then fitted to the circular polarization spectrum. As our
observations in V01 have shown that the observed V-spectrum can be
narrower than the synthetic V-spectrum, we also allow for the
narrowing of our synthetic spectra in the fit. Furthermore we need to
remove the scaled down replicas of the total intensity. We fit the
following function:

\begin{equation}
V_{\rm obs}(\Delta v) = a_{1}\cdot V^*(a_{2}\cdot \Delta v) +
a_{3}\cdot I(\Delta v)
\label{eq4}
\end{equation}

Here $\Delta v$ is the velocity in [km/s] measured from the peak of
the total intensity, $V^*$ is the synthetic spectrum $V_{\rm
synth}$. We fit for the parameters $a_1$, $a_2$ and $a_3$. The
parameter $a_1$ controls the amplitude of the V-spectrum, while $a_2$
determines the amount of narrowing of the observed spectra with
respect to the model spectra. Finally, the removal of the scaled down
total intensity profile is determined by $a_3$. The amplitude $a_1$ of
the best fitted model determines $V_{\rm min}$ and $V_{\rm
max}$. Together with the peak intensity of the maser feature this
gives $P_{\rm V}$. With the previously determined line width $\Delta
v_{\rm L}$ we calculate $B_{||}$ using Eq.~\ref{eq2}.

\subsubsection{Non-LTE}

 For the non-LTE case, we fit out models to both total intensity and
circular polarization simultaneously. We use Eq.~\ref{eq4} with $V^* =
V_{\rm model}$. Because the non-LTE models are already intrinsically
more narrow, we first fix the width of the V-spectrum by setting $a_2
= 1.0$ and only fitting $a_1$ and $a_3$. We have chosen to fit models
with $v_{\rm thermal} = 0.8$ and $1.0$ km/s, as the line widths of the
maser features indicate that much higher intrinsic thermal widths are
unlikely. We find that in some cases the circular polarization
spectrum is still narrower than the model spectrum. For these features
we allow for narrowing by releasing $a_2$. The combined fitting
determines the circular polarization percentage $P_{\rm V}$ and the
line width $\Delta v_{\rm L}$. The coefficient $A_{\rm F-F'}$ is
specifically determined for the best fitted model. Using this
Eq.~\ref{eq2} again gives $B_{||}$.

\section{Observations}
\label{obs}
\begin{table*}
\caption{Star Sample}
\begin{tabular}{|l|c|cc|c|c|c|c|}
\hline
Name & Type & RA (J2000) & Dec (J2000) & Distance & Period & V$_{\rm rad}$ & Peak
flux\\
&&($^{h}~^{m}~^{s}$)&($^{\circ}~{'}~{"}$)&(pc)&(days)&(km/s)&(Jy)\\
\hline
\hline
S Per & Supergiant & 02 22 51.72 & +58 35 11.4 & 1610$^a$ & 822 & -38.1 & 76.1\\
U Her & Mira & 16 25 47.4713 & +18 53 32.867 & 189$^b$ & 406 & -14.5 & 12.5
\\
NML Cyg & Supergiant & 20 46 25.7 & +40 06 56 & 1220$^c$ & 940 &
0.0 & 48.2\\
VY CMa & Supergiant & 07 22 58.3315 & -25 46 03.174 & 1500$^d$ & 2000 &
22.0 & 244.1\\
\hline
\multicolumn{8}{l}{$^a$ Hipparcos,$^b$van Langevelde et al. (2000),
  $^c$Dance et al.(2001) , $^d$Lada \& Reid (1978)}\\
\end{tabular}
\label{sample}
\end{table*}

The observations were performed at the NRAO\footnote{The National
Radio Astronomy Observatory is a facility of the National Science
Foundation operated under cooperative agreement by Associated
Universities, Inc.}  Very Long baseline Array (VLBA) on December 13th
1998.  At the frequency of the $6_{16}-5_{23}$ rotational transition
of H$_{2}$O, 22.235 GHz, the average beam width is $\approx 0.5 \times
0.5$ mas. This allows us to resolve the different H$_{2}$O maser
features in the CSE.  The data were correlated twice, once with modest
($7.8$ kHz $= 0.1$ km~s$^{-1}$) spectral resolution, which enabled us
to generate all 4 polarization combinations (RR, LL, RL and LR). The
second correlator run was performed with high spectral resolution
($1.95$ kHz $= 0.027$ km~s$^{-1}$), necessary to detect the circular
polarization signature of the H$_{2}$O Zeeman splitting, and therefore
only contained the two polarization combinations RR and LL.  We have
performed 6 hours of observations per source-calibrator pair. The
calibrator was observed for $1~1/2$ hours in a number of scans equally
distributed over the 6 hours. We used 2 filters (IFs) of 1 MHz width, which
were overlapped to get a velocity coverage of $\approx 22$ km/s.
This covers most of the velocity range of the H$_2$O maser.

\subsection{Calibration of the Data}
\begin{figure} 
   \resizebox{\hsize}{!}{\includegraphics{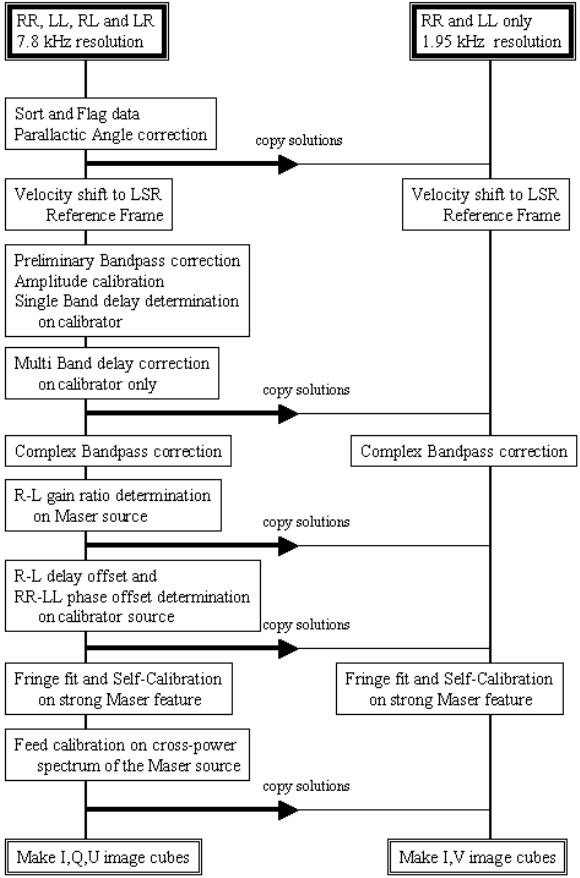}}
   \hfill
   \caption{The schematic data reduction path for our spectral line
   polarization data}  \label{fig1}
\end{figure}
 The data analysis followed the method of Kemball, Diamond \& Cotton
 (1995). The reduction path, shown in Fig.~\ref{fig1}, was performed
 in the AIPS data reduction package. The first standard calibration
 steps were performed on the data-set with modest spectral
 resolution. We used the system temperature measurements provided with
 the data to perform the amplitude calibration for both the
 calibrators and the sources. Parallactic angle correction, flagging
 and single- and multi-band delay calibration were all done regularly
 on the calibrators observed with each source. During 1/3 of the
 observation time, the first IF suffered strong interference, which
 forced us to flag several of the frequency channels ($\approx 0.1$
 MHz) in addition to the normal flagging. Also, for most of the
 observations IF 1 of the Los Alamos (LA) antenna was unusable.  The
 solutions obtained at these calibration steps were copied and applied
 to the high spectral resolution data set. The complex bandpasses were
 then determined for both data-sets separately. Additional calibration
 steps were needed for accurate processing of polarization data.  The
 gain ratio between the R- (right-) and L- (left-) hand polarizations
 was determined using the auto-correlation data of the reference
 antenna on a short scan of the maser source. This step contains the
 critical assumption of equal flux for the R- and L-handed circular
 polarizations. As a result the circular polarization spectra are
 forced to be anti-symmetric, as discussed below. Later, during the
 analysis of the circular polarization spectra, we still have to
 correct for small replicas of the total power profiles due to
 remaining small gain differences. After the gain calibration, we used
 the calibrators to determine the R-L delay offset and the RR-LL phase
 offset. The delay offset is expected to be stable over the duration
 of the observation (Brown et al. 1989) and can be determined from a
 single continuum scan in which the cross-hand fringes RL and LR are
 well detected. The RR-LL phase offset varies over the observation and
 are determined from calibrators which are assumed to show no circular
 polarization. The solutions were again copied from the modest
 resolution date and applied to the high resolution data. Then fringe
 fitting and self-calibration for the two separate data-sets were
 performed on a strong maser feature. Finally, corrections were made
 for the instrumental feed polarization using a range of frequency
 channels on the maser source, in which the expected linear
 polarization or the frequency averaged sum of the linear polarization
 is low. After the solutions were applied to both data sets image
 cubes could be created.

\subsection{Sample}

We observed 4 late type stars, the supergiants S~Per, VY~CMa and
NML~Cyg and the Mira variable star U~Her. They are listed in
Table.~\ref{sample} with type, position, distance, period and
velocity. The sources were selected on the basis of 2 criteria; strong
H$_2$O masers and previous SiO and OH maser polarization observations.
The peak fluxes measured in these observations are also shown in
Table.~\ref{sample}. On the total intensity channel maps with high
spectral resolution, the noise is dominated by dynamic range effects
and lies between $\approx 0.08 - 0.3$~Jy. On the circular polarization
maps we have noise of $\approx 0.01 - 0.03$~Jy.

The circular polarization of the SiO masers around VY~CMa has
previously been observed with a single dish by Barvainis et
al. (1987).  They find circular polarization of $6.5~\%$, indicating a
field strength of $B_{\rm SiO}\approx 65$~G. Observations of the 1612
MHz main line OH maser by Cohen et al. (1987) give $B_{\rm OH}\approx
2$ mG. Smith et al.(2001) concluded from ground based and Hubble Space
Telescope optical observation, that high magnetic fields of at least
$1$~G are necessary to explain the outflow of matter observed in
VY~CMa.  Cohen et al. also detected fields of $\approx 2$ mG on the
1612 MHz OH masers around NML Cyg, while Masheder et al. (1999)
estimated a field of $\approx 1$ mG on the OH masers around S~Per.  On
U~Her, Palen \& Fix (2000) observed pairs of R- and L-polarized maser
features, for the 1665 and 1667 MHz OH masers. Although not many of
these Zeeman pairs were found they estimate a magnetic field of
$B_{\rm OH}\approx 1$ mG.

\section{Results}
\begin{figure*} 
   \resizebox{\hsize}{!}{\includegraphics{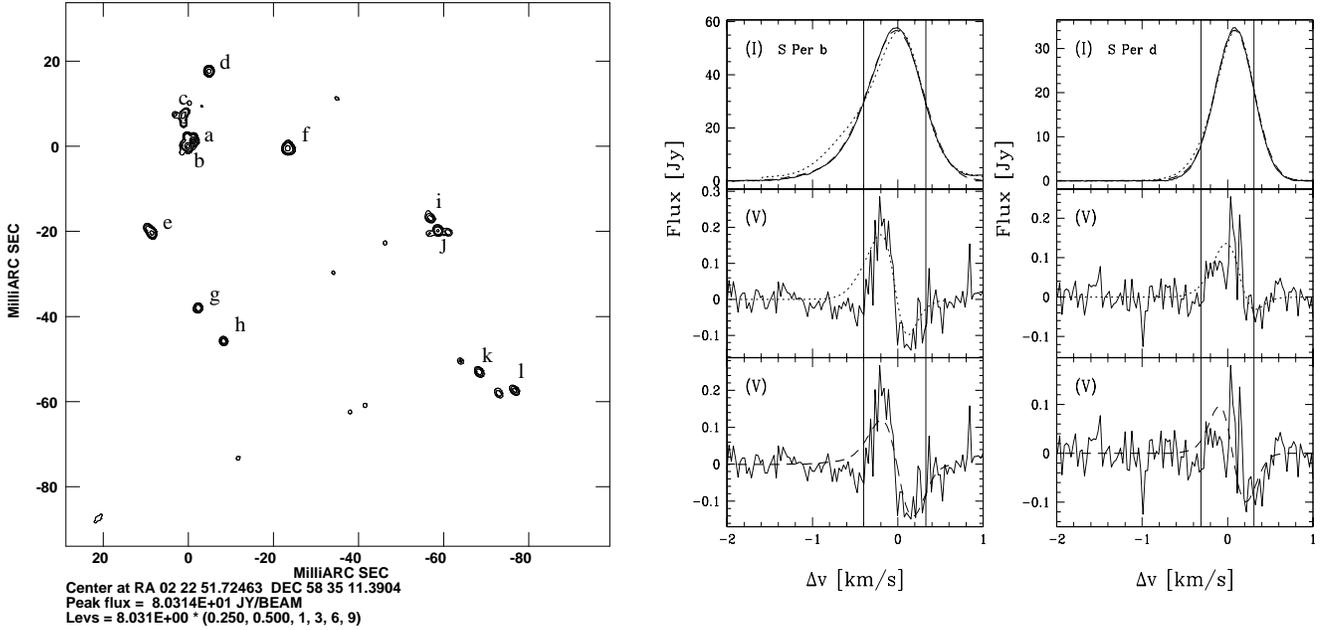}} \hfill
   \caption{(left) Total intensity image of the H$_{2}$O maser
   features around S~Per (right). Total power (I) and V-spectra for
   selected maser features of S~Per. The bottom panel shows the best
   fitting synthetic V-spectrum produced by the standard Zeeman
   interpretation (dashed line). The middle panel shows the best
   non-LTE model fit (dotted line). The corresponding total power fits
   are shown in the top panel. The V-spectra in the lower two panels
   are adjusted by removing a scaled down version of the total power
   spectrum determined in the fit by the $a_3$ parameter in
   Eq.\ref{eq4}, which is different for the LTE and non-LTE fits. The
   solid vertical lines show the expected position of minimum and
   maximum of the sine-shaped function in the general Zeeman case.}
\label{sper}
\end{figure*}

\begin{table*}
\caption{Results}
\begin{tabular}{|l|c|c|c|c|c|c|c|}
\hline
Name & Feature  & Flux (I)~(Jy) & $V_{\rm rad} {\rm (km/s)}$ & $\Delta
v_{\rm L} ~{\rm (km/s)}$ & P$_{\rm V}~(\cdot10^{-3})$ &
B~${\rm cos}\theta~{\rm (mG)}$ LTE& B~${\rm cos}\theta~{\rm (mG)}$ non-LTE\\
\hline
\hline
S Per & a & 76.1 & -27.2 & 0.88 & & & $<$ 26\\  
 & b & 57.7 & -30.9 & 0.77 & 6.20 & -207$\pm$30 & -132$\pm$18 \\
 & c & 11.4 & -33.9 & 1.21 & & & $<$ 254 \\
 & d & 34.7 & -37.1 & 0.54 & 5.62 & -140$\pm$32 & -77$\pm$17 \\
 & e & 29.7 & -28.9 & 0.61 & & & $<$ 41 \\
 & f & 35.6 & -40.1 & 0.53 & & & $<$ 33 \\
 & g & 15.4 & -30.4 & 0.92 & & & $<$ 125\\
 & h & 14.9 & -34.2 & 0.48 & & & $<$ 75 \\
 & i & 12.9 & -39.3 & 0.69 & & & $<$ 121 \\
 & j$^*$ & 26.6 & -39.2 & 0.64 & 2.41 & -156$\pm$47 & -68$\pm$35 \\
 & k & 14.2 & -27.2 & 0.51 & & & $<$ 72 \\
 & l & 10.0 & -23.2 & 0.44 & & & $<$ 110 \\
\hline
VY CMa & a$^*$ & 244.1 & 10.8 & 0.66 & 3.70 & -116$\pm$21 & -151$\pm$21 \\
 & b & 155.5 & 10.6 & 0.67 & 5.64 & 182$\pm$30 & 99$\pm$15 \\
 & c & 115.2 & 12.9 & 0.74 & 2.73 & 85$\pm$18 & 112$\pm$19 \\
 & d & 47.9 & 13.7 & 0.83 & 3.24 & 129$\pm$31 & 81$\pm$16 \\
 & e & 17.8 & 15.0 & 0.94 & 4.10 & 164$\pm$66 & 166$\pm$55 \\
 & f & 58.5 & 18.3 & 0.80 & & & $<$ 25 \\
 & g & 25.3 & 27.1 & 0.88 & & & $<$ 55 \\
 & h & 43.1 & 28.8 & 0.61 & & & $<$ 25 \\
 & i & 65.6 & 25.2 & 0.79 & 2.23 & 83$\pm$25 & 48$\pm$14 \\
 & j & 126.2 & 24.6 & 0.60 & 4.66 & 133$\pm$26 & 85$\pm$12 \\
\hline
NML Cyg & a$^*$  & 48.2 & -21.2 & 0.66 & 5.9 & 215$\pm$37 & 118$\pm$15 \\
 & b & 10.5 & -20.4 & 0.40 & 13.8 & 234$\pm$63 & 221$\pm$41 \\
 & c & 5.42 & -20.2 & 0.54 & 19.2 &  492$\pm$133 &  336$\pm$67 \\
 & d & 3.66 & -20.0 & 0.54 & 21.6 &  553$\pm$220 &  428$\pm$149 \\
\hline
U Her &  a & 12.20 & -17.8 & 0.75 & 27.5 & 858$\pm$161 & 605$\pm$83 \\
 & b & 2.91 & -17.6 & 0.50 & 126.1 & 2434$\pm$505 & 1487$\pm$231 \\
 & c & 1.16 & -17.6 & 0.73 & & & $<$ 803 \\
 & d$^*$ & 2.46 & -19.2 & 0.47 & 49.6 & 1181$\pm$297 & 552$\pm$114 \\
 & e$^*$ & 2.07 & -19.2 & 0.51 & 41.1 & 980 $\pm$298 & 726$\pm$189 \\
 & f & 1.36 & -19.3 & 0.51 & & & $<$ 337 \\
\hline
\multicolumn{8}{l}{$^*$ See text}
\end{tabular}
\label{results}
\end{table*}

We have examined the strongest H$_2$O maser features around the 4
stars observed. The results are shown in Table~\ref{results} for
features with fluxes up to $10\%$ of the brightest maser spots. Some
selected spectra are shown with fits of the LTE Zeeman models
and the non-LTE models. When using the non-LTE models, fits with
$v_{\rm thermal} = 1.0$ km/s generally are best. As expected the
splitting is small and the circular polarization is generally not more
than $0.5 - 1.0\%$. Magnetic field strengths obtained using the LTE
Zeeman models are given in column 7 while the results obtained
using the non-LTE models are shown in column 8. The features that show
less than a $3\sigma$ circular polarization signal are considered
non-detections.  For these we have determined (absolute) upper limits
using the $A_{\rm F-F'}$ coefficients obtained using the NW radiative
transfer models.  When using the LTE Zeeman analysis the limits
increase by a factor of $\approx 1.4$.

 Features labeled $^*$ have V-spectra that did not allow accurate
fitting, possibly due to blending effects. They do show circular
polarization above the $3\sigma$ level, although generally only
slightly more. For these we have used $V_{\rm min}$ and $V_{\rm max}$
to estimate the magnetic field strength, and we assumed the best
estimate for the coefficient $A_{\rm F-F'} = 0.018$. The errors are the
formal $1\sigma$ errors.

 We have not been able to detect linear polarization in any of the 4
sources. This gives upper limits to the fractional linear polarization
of $\approx 1\%$ for the weakest features down to $\approx 0.01\%$ for
the strongest.

\subsection{S~Per}

Fig.~\ref{sper} shows the total intensity map of the H$_2$O maser
features around S~Per, in which we can identify most of the features
detected in earlier observations(Diamond et al. 1987; Marvel 1997).
Positions are shown relative to the feature labeled {\it S~Per~b},
discussed in more detail in V01. Circular polarization between $0.2 -
0.7\%$ has been detected in 3 of the brightest features after a
careful reexamination of the data. This has also resulted in a
slightly lower magnetic field strength value for {\it S~Per~b} than
the one presented in V01 after optimizing the fitting routines. The
figure also shows the total power and circular polarization spectra
for 2 of the features, with fits for both the LTE and non-LTE Zeeman
models. The velocity labeling has changed from V01 to
correspond to the convention used by NW. We find the magnetic field
pointing away from us on all features. From these results we estimate
the magnetic field in the H$_2$O maser region around S~Per to be
$\approx$ 200 mG using the LTE Zeeman analysis or $\approx$ 150
mG using the non-LTE models.

The maximum radius of the H$_2$O maser region was determined to be
$\approx 60$ mas by Diamond et al.(1987), corresponding to $\approx
100$ AU. They conclude for a thick shell model that the inner radius
of the water maser shell is $\approx 30$ mas, while the outer radius
is $\approx 70$ mas. For comparison, the OH maser shell extent was
determined by them to be $80 \times 30$ mas, using MERLIN
observations.
    
\subsection{VY~CMa}
\begin{figure*} 
   \resizebox{\hsize}{!}{\includegraphics{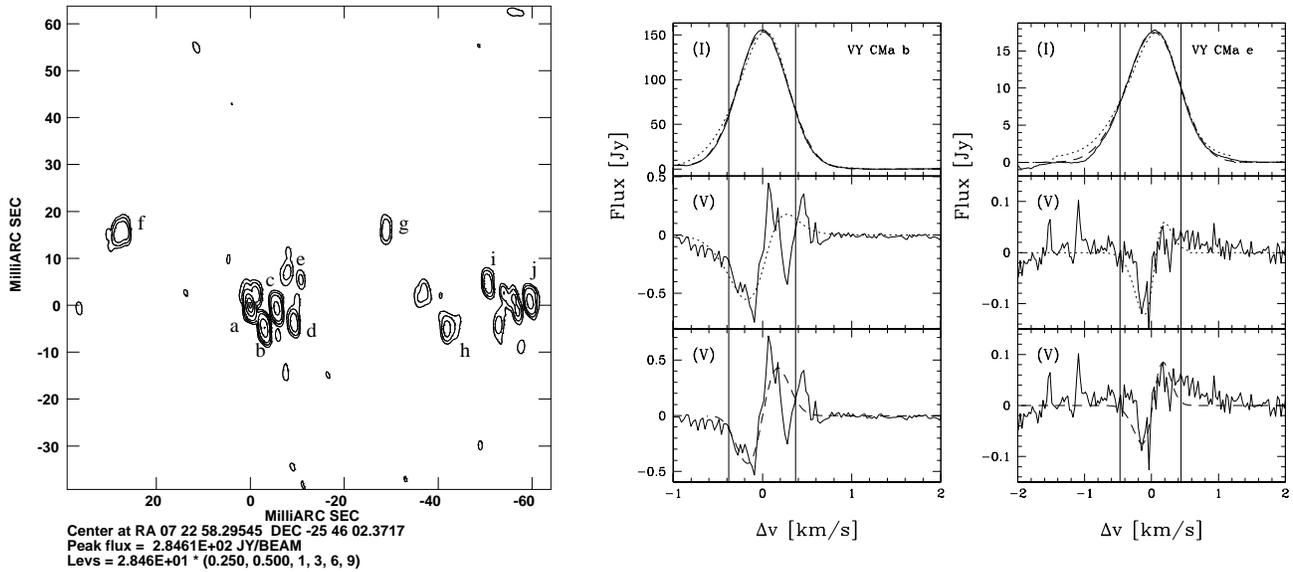}}
   \hfill \caption{Similar to Fig.~\ref{sper} for VY~CMa}  \label{vycma}
\end{figure*}

 The total intensity map of the H$_2$O masers around VY~CMa is shown
in Fig.~\ref{vycma}.  Circular polarization has been detected on 7 out
of 10 of the brightest maser features down to $P_{\rm V}\approx 0.2\%$. Similar
to the values found for S~Per we find a magnetic field of $\approx
200$~mG using the LTE Zeeman analysis or $\approx 175$ mG using the
NW models. The magnetic field points toward us for most of the
sources. {\it VY~CMa~a} suffers from blending with nearby maser spots
which made accurate fitting impossible, this can also explain the
change of sign.

Diamond et al.(1987) give the radius of the H$_2$O maser region for
VY~CMa to be $\approx 150$ mas, corresponding to $\approx 220$ AU. The
1612 MHz OH maser radius is shown to be $\approx 2400$ AU in
observations by Reid et al.(1981).

\subsection{NML~Cyg}
\begin{figure*} 
   \resizebox{\hsize}{!}{\includegraphics{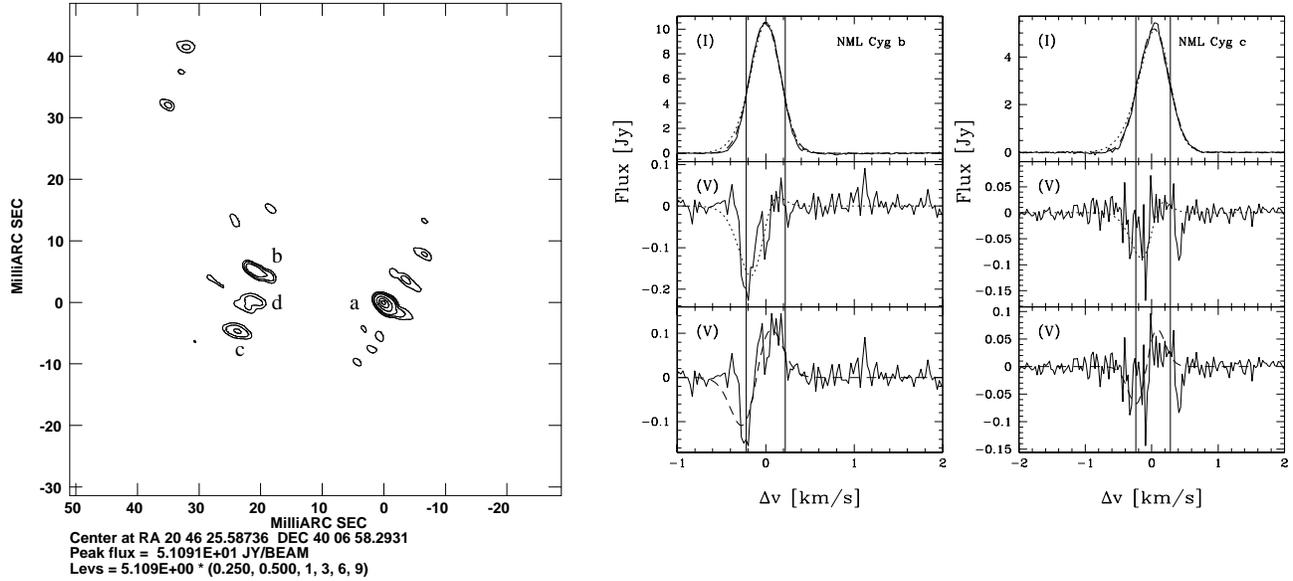}}
   \hfill \caption{Similar to Fig.~\ref{sper} for NML~Cyg}  \label{nmlcyg}
\end{figure*}

\begin{figure} 
   \resizebox{\hsize}{!}{\includegraphics{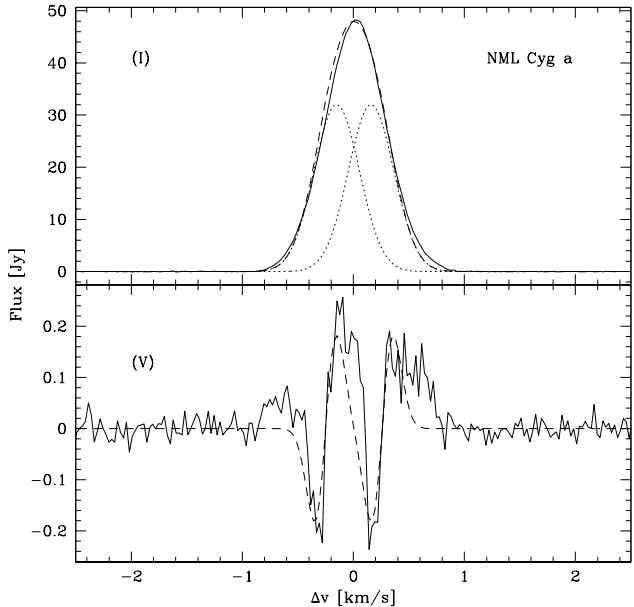}} \hfill
   \caption{Total power (I) and circular polarization (V) of NML Cyg
   a. Dotted and dashed lines are the model of two slightly shifted
   spatially blended maser features.}  \label{sp4}
\end{figure}

 We have only been able to detect a few H$_2$O maser features around
NML~Cyg, which are shown in Fig.~\ref{nmlcyg}. We find circular
polarization from $0.6 - 2\%$. The spectra for {\it NML~Cyg~b} and
{\it c} are shown in the same Figure. The V-spectrum for {\it
NML~Cyg~a} is shown in Fig.~\ref{sp4}. The circular polarization
spectrum seems to indicate that this features actually consists of 2
heavily blended features of $\approx 30$ Jy. The magnetic field
strength on this feature has been estimated using this simple model.
Both the interpretations indicate the magnetic field strength to be
$\approx 500$ mG, higher than for both S~Per and VY~CMa.

The H$_2$O maser region extent is $\approx 195$ mas, corresponding to
$\approx 240$ AU (Johnston et al. 1985). Diamond et al.(1984) give for
the 1612 MHz OH masers a maximum extent of $\approx 3-5$ arcsec. 

\subsection{U~Her}
\begin{figure*} 
   \resizebox{\hsize}{!}{\includegraphics{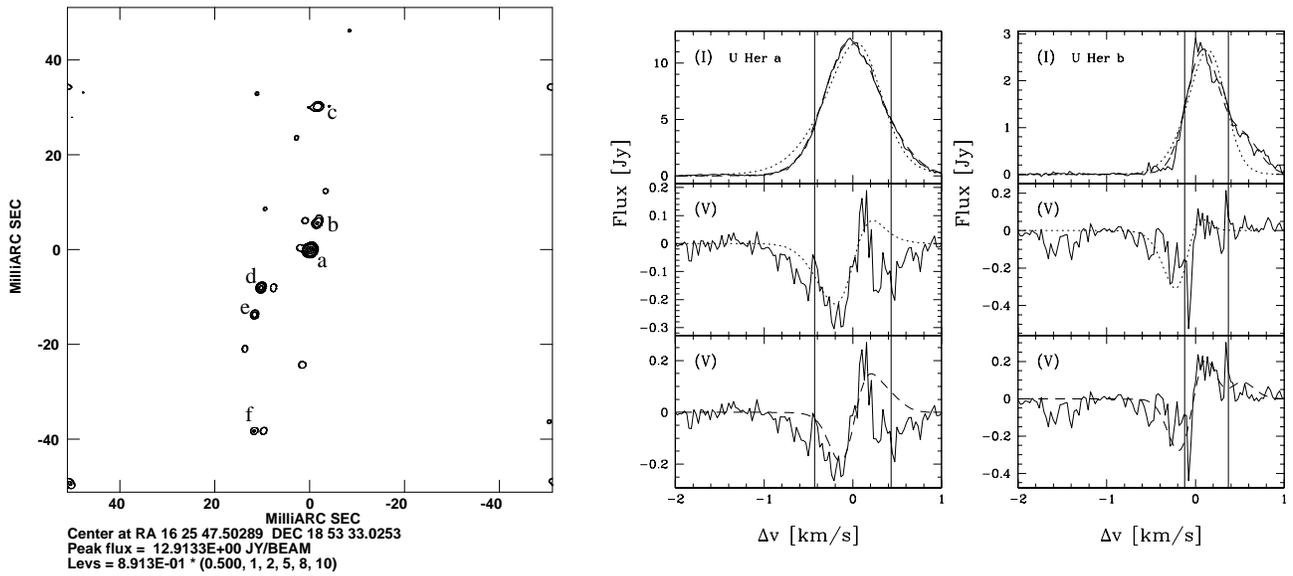}}
   \hfill \caption{Similar to Fig.~\ref{sper} for U~Her}  \label{uher}
\end{figure*}

 The total intensity map for the H$_2$O masers around U~Her is shown
in Fig.~\ref{uher}. U~Her is the only Mira variable star in our sample
and shows significantly weaker H$_2$O masers. However as seen in the
figure, we do detect circular polarization of up to $\approx
10\%$. The field strength in the H$_2$O maser region of U~Her is
higher than those observed for the supergiant stars in our sample. We
find strengths of $\approx 2.5$ G using the LTE Zeeman
analysis or $\approx 1.5$ G using the non-LTE
models. While the features {\it U~Her~a} and {\it b} show a clear
sine-shaped V-spectrum, the features {\it d} and {\it e} have circular
polarization signals only slightly outside the $3\sigma$ detection
limit. This made fitting actual profiles impossible for those
features.

The size of the H$_2$O maser region is found to be $\approx 350 \times
200$ mas (Bowers et al. 1989), which at 189 pc distance corresponds to
$\approx 65 \times 40$ AU.  Chapman et al. (1994) find for the OH
maser extent of U~Her of $800 \times 900$ mas, $\approx 150 \times
170$ AU.

\section{Discussion}

\subsection{Interpretation of the Polarization Results}

 The observations show magnetic fields in the range from $\approx 150$ mG to
$\approx 1.5$ G in the H$_2$O maser regions of our sample of stars. The
exact field strength depends on the interpretation used. Here we have
used the LTE Zeeman analysis presented in FG and V01, as
well as a non-LTE model presented by NW. Generally, the
results using the NW method give fields approximately $2/3$ of the
LTE Zeeman method. As shown above, the non-LTE
models do not typically produce anti-symmetric shapes for the circular
polarization spectra. However, the anti-symmetric shapes observed are a
necessary result of the data treatment, when it is assumed that the R-
and L-polarization profiles are similar and when the scaled down
replica of the total power spectrum is removed. We can thus fit both
interpretations to the data by leaving the amplitude of the total power
replicas as a free parameter as described in Section~\ref{sect}. 
As a consequence, we are unable to directly distinguish between the
different interpretations on the basis of the anti-symmetry of the
V-spectra. 

The shape and line widths of the total intensity spectra of the
different maser features indicate that for all stars the masers are
unsaturated or only slightly saturated. As a result, even for the non-LTE
case, we do not expect significant deviations from
anti-symmetry. The line widths also indicate that $v_{\rm thermal}
\approx 1.0$ km/s is indeed the best estimate for the thermal
line width. In the cases where we could not perform good fits to the
spectra, we have thus used $A_{\rm F-F'} = 0.018$ as the best estimate.
Since the masers are thought to be unsaturated the circular
polarization decreases linearly with ${\rm cos}~\theta$, as shown in
Fig.~\ref{theta}.

\begin{figure} 
   \resizebox{\hsize}{!}{\includegraphics{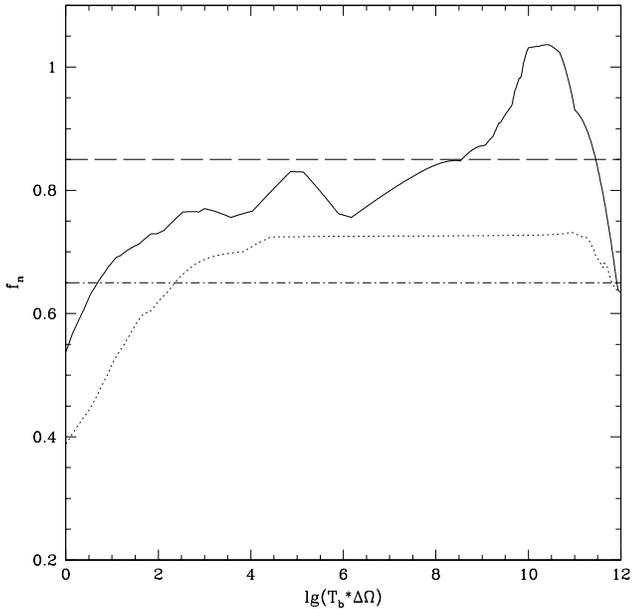}} \hfill
   \caption{Fractional narrowing of the non-LTE models
   as a function of emerging brightness temperature. The solid line
   indicates the uni-directional maser, the dotted line is the
   bi-directional maser. Both models have $v_{\rm thermal} = 1.0$
   km/s. The long-dashed line is the theoretical value for the LTE Zeeman
   analysis, while the dashed-dotted line is our measured average
   narrowing value.}  \label{frac}
\end{figure}

An important motivation for using the non-LTE approach, is the fact
that the circular polarization spectra show narrowing that is not
expected in the LTE case, as discussed in V0. The observed spectra
are not directly proportional to $I'$, as the minimum and maximum are
closer together than predicted. We can express this narrowing as
$f_{\rm n} = \Delta v_{\rm L} / \Delta v_{\rm mm}$, where $\Delta
v_{\rm L}$ is the FWHM of the maser feature and $\Delta v_{\rm mm}$
the separation between the minimum and maximum of the
V-spectrum. $\Delta v_{\rm mm}$ is determined after the spectrum is
forced to be anti-symmetric by fitting the scaled down total power
spectrum.  In our spectra we find on average $f_{\rm n} \approx 0.64$,
but sometimes as low as $0.45$, which cannot be explained by LTE
Zeeman models. In that case, when $V \propto I'$, we find
$f_{\rm n} \approx 0.85$. The non-LTE models partly explains this
narrowing. As a result of the narrowing and rebroadening of the maser
lines and of the hyperfine interaction, $f_{\rm n}$ depends on the
maser brightness as shown in Fig.~\ref{frac}. The narrowing fraction
decreases to $\approx 0.6$ in the case of highly saturated
masers. However, as discussed above, the maser are indicated to be
mostly unsaturated and thus the reason for the narrowing is still not
completely clear. We have performed some tests using bi-directional
masers. As seen in Fig.~\ref{frac}, these show more narrowing over the
entire range, while the limit for the saturated masers is similar to
the uni-directional masers. Thus, although the predicted narrowing
fraction is still higher than observed, this is probably due to the
assumption of linear masers in the models. Aside from the narrowing of
the circular polarization profiles for bi-directional masers, the
resulting $A_{\rm F-F'}$ coefficients and corresponding magnetic
fields do not change. We have used the uni-directional models for the
fits, since the computation time increased considerably when
calculating the bi-directional case.

The above indicates that the narrowing of the V-spectrum is an important
diagnostic for distinguishing which method yields the most reliable
magnetic field strengths. A careful analysis of the total power
intensity profiles will also be able to distinguish between different
interpretations. The fits presented for our sources show that the
non-LTE models do not fit the total intensity profiles as well as the
fit of 3 hyperfine components for the LTE Zeeman models.  However,
this is due to the fact that it was impossible to completely explore
the full parameter space for the models.  Fortunately, this has only a
small effect on the magnetic field strengths inferred by this method,
which has been included in the quoted errors. We thus conclude that
the best estimates for the magnetic field strengths in the H$_2$O
maser region are those obtained using the non-LTE
models.

We have been unable to detect any linear polarization for the sources
in our sample. The upper limits are between $0.01$ and $1.0 \%$.  This
is consistent with previous observations by Barvainis \& Deguchi
(1989), and is a separate indication that the masers are not strongly
saturated. The absence of linear polarization excludes the non-Zeeman
interpretation. We are therefore confident that the observed circular
polarization is the signature of the actual Zeeman splitting and that
our inferred values are the best estimates for the magnetic field.

\subsection{Magnetic fields in CSE}

We can compare our results with the magnetic fields measured on the
other maser species in the circumstellar envelope. As remarked before
there have been several observations to determine the magnetic fields
in the SiO maser regions (e.g. Barvainis et al. 1987, Kemball \&
Diamond 1997).  Assuming that the circular polarization observed on
these sources is caused by Zeeman splitting, fields between $50$ and
$100$ G have been inferred. This had already been predicted earlier
from observations of the Zeeman splitting in OH masers, which
indicated fields of $1-2$ mG in Mira stars and up to $\approx 10$ mG
in supergiants (Reid et al. 1979, Claussen \& Fix 1982). Various other
observations of OH masers have since then confirmed these milliGauss
fields. The high field strengths on the SiO masers however, remain the
topic of debate. It has been shown that fields of not more than
$\approx 30$ mG can cause the high circular polarization observed; as
described in the non-Zeeman effect above. Also, Elitzur (1996) argues
that the magnetic fields inferred on both the SiO and the OH maser
could be lower by a factor of $8$ -- $10$.

Additionally, fields of tens to hundreds of Gauss in the SiO maser
region would indicate fields of the order of $10^3$ G on the surface
of the star. It was argued that such fields could not be produced by
AGB stars (Soker \& Harpaz, 1992). The high magnetic fields were
determined by assuming the field strength varies with distance from
the star by the relation $B \propto r^{\alpha}$. The exponent $\alpha$
depends on the structure of the magnetic field in the circumstellar
envelope. A solar-type magnetic field has $\alpha = -2$, while for a
dipole medium $\alpha = -3$.

\begin{figure} 
  \resizebox{\hsize}{!}{\includegraphics{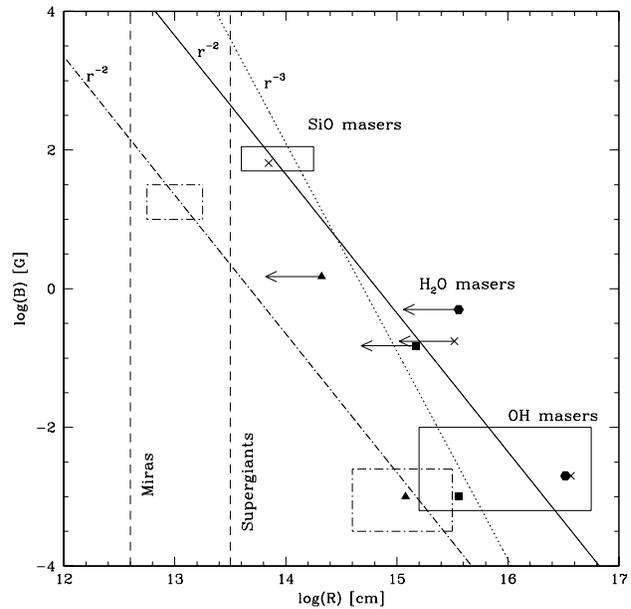}} \hfill
  \caption{Magnetic Field strength, B, as function of distance, R,
  from the center of the star. Dashed-dotted boxes are the estimates
  for the Mira variable stars, solid boxes are those for the
  supergiant stars. Similarly, the solid and the dashed-dotted line
  corresponds to a solar-type magnetic field for the supergiants and
  the Mira variable stars respectively. The dotted line indicated a
  dipole magnetic field for supergiant stars. Symbols indicate
  observations. U~Her is represented by triangles, S~Per by the
  square, VY~CMa by the crosses and NML~Cyg by the hexagonal
  symbols. The dashed lines represent an estimate of the stellar
  radius.}  \label{rb}
\end{figure}

For a comparison between the different maser species, the results from
this paper are shown in Fig.~\ref{rb}, combined with results obtained
on our sample of stars for the other masers. The figure displays the
magnetic field strength as a function of distance from the star.  We
have indicated the actual measurements for the magnetic field with
different symbols for each star. The boxes indicate estimates for both
magnetic fields and distance of the maser species based on
observations of different stars (Johnston et al. 1985; Lane et al.
1987; Barvainis et al. 1987; Lewis et al. 1998). We have separated
the Mira variable stars and the supergiants. A major uncertainty is
the distance of the maser features for which the magnetic fields were
determined. The SiO masers are expected to exist in a narrow region at
approximately 1 stellar radius from the star (Reid \& Menten 1997). The
radial distances are obtained from the observation of their
characteristic ring structure (e.g. Diamond et al. 1994).  Several
observations have determined the extent of the OH maser regions of our
sources (Diamond et al. 1987, Chapman \& Cohen 1986).  However,
because the OH maser is strongly radially beamed these distances are
only lower limits to the actual extent, which for the supergiant stars
could be a few times $10^{16}$ cm. The plotted measurements use the
literature values quoted above as the value of the radial
distance. The extent of the H$_2$O maser region has also been the
object of several studies (e.g. Bowers et al. 1989, Lane et
al. 1987). For our observations we have used the values quoted
above. As the figure shows, our observations lie systematically above
both the solar-type and the dipole magnetic field approximation.  This
can be caused by a selection effect. Our observations are heavily
biased for the highest magnetic fields, which occur closer to the
star. Thus, while the distance values used in the plot are the outer
edges of the H$_2$O maser region, we most probably probe the inner
regions. So using the solar-type magnetic field, we can estimate the
thickness of the H$_2$O maser region. This ranges from $1 - 2 \cdot
10^{15}$ cm for the supergiant stars to $\approx 2\cdot 10^{14}$ cm
for U~Her. This agrees with values found using the pumping models for
H$_2$O masers in late-type stars presented by Cooke \& Elitzur (1985),
and also with the estimates presented for S~Per in Diamond et
al.(1987).

Another possibility is the existence of high density H$_2$O maser
clumps with frozen in magnetic field lines. For frozen magnetic fields
the field strength varies with number density as $B \propto n^k$, with
$1/3 \leq k \leq 1/2$ from theoretical predictions (e.g. Mouschovias,
1987), where $n$ is the number density. Density ratios between the
maser clumps and their surrounding medium need to be between $\approx
10$ and $\approx 1000$ to explain the observations if we assume
that the magnetic field strength in the medium surrounding the clumps
is similar to the field strength observed on the OH masers. While the
higher values are unlikely, the lower values are certainly
reasonable. Because the observations are most sensitive for the high
magnetic fields, we are in this interpretation biased toward the
highest density maser clumps. A combination of the two effects
described is likely.

The pressure of the magnetic fields measured in the H$_2$O maser
region dominate the thermal pressure of the gas. The ratio between
thermal and magnetic pressure is given by $\beta \equiv (8\pi n_{\rm
H} k T)/B^2$. Here $k$ is the Boltzman constant. Assuming a gas
density of $n_{\rm H} = 10^9$ cm$^{-3}$ and a temperature of $T\approx
10^3$ K at the inner edge of the H$_2$O maser region, and taking
$B \approx 250 $ mG, $\beta \approx 0.05$, indicating that the
magnetic pressure dominates the thermal pressure by a factor $\approx
20$.

The inferred values for the magnetic field strength on the H$_2$O
masers are intermediate between the values determined on the OH masers
and those determined for the SiO masers with the standard Zeeman
interpretation. This can be viewed as evidence that the measured
circular polarization on the SiO masers is caused by actual Zeeman
splitting and that the non-Zeeman interpretation is unlikely even
though the SiO masers exhibit up to $100 \%$ linear polarization.  Our
observations agree with a power-law dependence of the magnetic field
on distance to the star, of which the solar-type field seems most
likely. An order of magnitude estimate for the magnetic fields on the
surface of the star then gives fields of $\approx 10^2$ G for Mira
variable stars and $\approx 10^3$ for supergiants. Recently, Blackman
et al. (2001) have shown that an AGB star can indeed produce such
strong magnetic fields, originating from a dynamo at the interface
between the rapidly rotating core and the more slowly rotating stellar
envelope.
  
\section{Conclusions}

The analysis of the circular polarization of H$_2$O masers depends
strongly on the models used. The interpretation of our observations
has shown that the non-LTE, full radiative transfer models presented
by NW improve the accuracy of the fits. Some aspects of the
observations however, remain unclear, even though the non-LTE approach
can explain most of the narrowing of the circular polarization
spectra. Multi-dimensional models seem promising to explain the
narrowing observed. We have found magnetic field strengths of $\approx
150-500$ mG for the supergiants and $\approx 1.5$ G for the Mira
variable U~Her. They occur in the most dense H$_2$O maser spots at the
inner boundaries of the maser region and indicate strong magnetic
fields at the surface of the star. The fields dominate the thermal
pressure of the gas and are strong enough the drive the stellar
outflows and shape the stellar winds.

{\it Acknowledgments:} 
This project is supported by NWO grant 614-21-007.


\begin{thebibliography}{999}




\bibitem[1991]{And}
Anderson, N., Watson, W.D., 1993, ApJ, 407, 620.

\bibitem[1987]{Barv}
Barvainis, R., McIntosh, G., Predmore, C.R., 1987, Nature, 329, 613.

\bibitem[1989]{BD}
Barvainis, R., Deguchi, S., 1989, AJ, 97, 1089.

\bibitem[2001]{Black}
Blackman, E.G., Frank, A., Markiel, J.A., et al., 2001, Nature, 409, 585.

\bibitem[1989]{Bow}
Bowers, P.F., Johnston, K.J., de~Vegt, C., 1989, ApJ, 340, 479. 

\bibitem[1989]{Brown}
Brown, L.F., Roberts, D.H., Wardle, J.F.C., 1989, AJ, 97, 1522.

\bibitem[1986]{Chap2}
Chapman, J.M., Cohen, R.J., 1986, MNRAS, 220, 513.

\bibitem[1994]{Chap}
Chapman, J.M., Sivagnanam., P., Cohen, R.J., Le~Squeren, A.M., 1994,
MNRAS, 268, 475.

\bibitem[1982]{Claus}
Claussen, M.J., Fix, J.D., 1982, ApJ, 263, 153.

\bibitem[1987]{cohen}
Cohen, R.J., Downs, G., Emerson, R., et al., 1987, MNRAS, 225, 491.

\bibitem[1985]{cook}
Cooke, B., Elitzur, M., 1985, ApJ, 295, 175.

\bibitem[2001]{dance}
Dance, W.C., Green, W.H., Hale, D.D.S, et al., 2001, ApJ, 555, 405.

\bibitem[1986]{DW}
Deguchi, S., Watson, W.D., 1986, ApJ, 302, 750.

\bibitem[1984]{Diam2}
Diamond, P.J., Norris, R.P., Booth, R.S., 1984, MNRAS, 207, 611.

\bibitem[1987]{Diam}
Diamond, P.J., Johnston, K.J., Chapman, J.M., et al., 1987, AA, 174, 95.

\bibitem[1994]{Diam3}
Diamond, P.J., Kemball, A.J., Junor, W., et al., 1994, ApJ, 430, L61.

\bibitem[1996]{Elitz}
Elitzur, M., 1996, ApJ, 457, 415.

\bibitem[1989]{FG}
Fiebig, D., G\"usten, R., 1989, AA, 214, 333. (FG)

\bibitem[1985]{John}
Johnston, K.J., Spencer, J.H., Bowers, P.F., ApJ, 290, 660.

\bibitem[1995]{KDC}
Kemball, A.J., Diamond, P.J., Cotton, W.D., 1995, A\&AS, 110, 383.

\bibitem[1997]{KD}
Kemball, A.J., Diamond, P.J., 1997, ApJ, 481, L111.

\bibitem[1978]{lada}
Lada, C.J., Reid, M.J., 1978, ApJ, 219, 95.

\bibitem[1987]{lane}
Lane, A.P., Johnston, K.J., Bowers, P.F., et al., 1987, ApJ, 323, 756.

\bibitem[1998]{lewis}
Lewis, B.M, 1998, ApJ, 508, 831.

\bibitem[1997]{Marvel}
Marvel, K., 1997, Ph.D Thesis, new Mexico State University.

\bibitem[1999]{Mash} 
Masheder, M.R.W., Van~Langevelde, H.J., Richards, A.M.S., et al.,
1999, NewAR, 43, 563.

\bibitem[1987]{Mous}
Mouschovias, T.Ch, 1987, Physical Processes in Interstellar Clouds,
eds. G.E. Morfill, M. Scholer, Reidel, Dordrecht, p.453.

\bibitem[1990]{NW1}
Nedoluha, G.E., Watson, W.D., 1990, ApJ, 361, L53. 

\bibitem[1991]{NW1a}
Nedoluha, G.E., Watson, W.D., 1991, ApJ, 367, L63. 

\bibitem[1992]{NW}
Nedoluha, G.E., Watson, W.D., 1992, ApJ, 384, 185. (NW)

\bibitem[1990]{NM}
Neufeld, D.A., Melnick, G.J., 1990, ApJ, 352, L9. 

\bibitem[2000]{pf}
Palen, S., Fix, J.D., 2000, ApJ, 531, 391.

\bibitem[1979]{Reid2}
Reid., M.J., Moran, J.M., Leach, R.W., et al., 1979, ApJ, 227, L89. 

\bibitem[1981]{Reid}
Reid., M.J., Moran, J.M., Johnston, K.J., 1981, AJ, 86, 897. 

\bibitem[1997]{Reid3}
Reid., M.J., Menten, K.M., 1997, ApJ, 476, 327. 

\bibitem[1999]{Rich}
Richards, A.M.S., Yates, J.A., Cohen, R.J., 1999, MNRAS, 306, 954.

\bibitem[1992]{SH}
Soker, N., Harpaz, A., 1992, JASP, 104, 923. 

\bibitem[2001]{smith}
Smith, N., Humphreys, R.M., Davidson., K., et al., 2001, AJ, 121, 1111.

\bibitem[1997]{SzyCoh}
Szymczak, M., Cohen, R.J., 1997, MNRAS, 288, 945.

\bibitem[2000]{vL}
van~Langevelde, H.J., Vlemmings, W., Diamond, P.J., et al., 2000, AA,
357, 945.

\bibitem[2001]{V01}
Vlemmings, W., Diamond, P.J., van Langevelde, H.J., 2001, AA, 375,
L1. (V01)

\bibitem[2001]{WW}
Watson, W.D., 2001, ApJ, 558, L55.

\bibitem[1998]{Wiebe}
Wiebe, D.S., Watson, W.D., 1998, ApJ, 503, L71.

\end{thebibliography}
\end{document}